\documentclass[aps,prl,twocolumn,groupedaddress,showpacs]{revtex4}
\usepackage{graphicx}% Include figure files
\usepackage{dcolumn}% Align table columns on decimal point
\usepackage{bm}% bold math

\begin{document}

\title{A prediction of string theory testable at low energies}

\author{Hrvoje Nikoli\'c}
\affiliation{Theoretical Physics Division, Rudjer Bo\v{s}kovi\'{c}
Institute,
P.O.B. 180, HR-10002 Zagreb, Croatia.}
\email{hrvoje@thphys.irb.hr}

\date{\today}

\begin{abstract}
Superstring theory involves a single unifying superstring current for bosons and fermions
that in the low-energy limit generalizes the Klein-Gordon (not the Dirac) current.
By adopting a relativistic-covariant probabilistic interpretation of the current,
the low-energy limit implies electron interference patterns that, under certain conditions,
differ from those predicted by the Dirac current.
\end{abstract}

%Strings and branes, String and brane phenomenology 

\maketitle
 
From the phenomenological point of view, the main problem with string theory 
\cite{GSW,polc,BBS} is that it is difficult to test this theory experimentally.
Typical generic predictions of string theory that differ from those of the
Standard Model occur at Planck energies, which cannot be reached experimentally 
with present and near-future technologies. The theory also allows phenomena, such as 
large extra dimensions \cite{dvali}, that are testable at energies much lower
than the Planck energy, but such phenomena are typically not generic predictions
of string theory, but correspond to very special solutions for which, {\it a priori},
there is no special reason for realization in our visible part of the Universe.
Even supersymmetry can, in principle, be broken at an arbitrarily large
scale, implying that supersymmetric particles do not necessarily need to be
seen at energies available experimentally. Thus,
a true progress in string phenomenology would be achieved
by finding a prediction that is both generic (i.e., does not correspond
only to a special solution) and testable at energies much lower
than the Planck energy. Any proposal for such a phenomenon
should be strongly wellcome.

In this paper we propose such a generic prediction of string theory testable
at energies much lower than the Planck energy. Specifically, we propose
a test of the distinguished property of string theory that 
the c-number valued wave functions of
bosonic and fermionic particles
are different states of the {\em same} object. We stress (i) that this property is a generic
property of all superstring theories, (ii) that it cannot be derived from 
low-energy supersymmetric field theory, and (iii) that, as we shall see,
it can be tested at low energies independently of the scale of supersymmetry breaking.

The idea can be explained shortly as follows. All physical string states satisfy
the Hamiltonian constraint which, in the Schr\"odinger picture, can be viewed
as a stringy generalization of the Klein-Gordon equation \cite{nikbosfer}.
With this equation one can also associate the corresponding conserved
superstring current generalizing the Klein-Gordon (not the Dirac) point-particle
current. Thus, whatever the physical interpretation of this current is,
this interpretation must be equally valid for both bosons and fermions.
Recently, a progress in the probabilistic interpretation of the Klein-Gordon current and its generalizations has been achieved \cite{nikijmpa,nikprob}, showing
that the absolute value of the time-component of the current can be consistently 
interpreted as the relativistic-covariant probability density of particle (or string) positions.
Curiously, this progress has been influenced by the Bohmian interpretation of
relativistic quantum mechanics (QM) according to which particles and strings move 
along integral curves 
of the currents \cite{durr96,tumul,nikfpl1,nikfpl3,nikstr1,nikstr2,nikbosfer,niknofield,nikprob}, 
but it turns out that these integral curves play an important role even within the
conventional probabilistic interpretation of relativistic QM \cite{nikijmpa}.
(For the role of Bohmian trajectories in the conventional interpretation of nonrelativistic QM, 
see also \cite{lopr}.) 
By applying this probabilistic interpretation to fermions (e.g., electrons), one obtains
a probability density that, in general, differs from the probability density obtained
from the Dirac current. Since without string theory one naturally expects
that the probability density of fermions should be given by the Dirac current,
a measurement of such a probability density can distinguish between 
a prediction of string theory and that of the conventional point-particle theory.

Before proceeding with the formal analysis, we also stress that the possibility
that the probability density of electron positions could be given by the 
Klein-Gordon current is not in contradiction with the fact that
the electron field is coupled to gauge fields via the Dirac current.
The probabilistic interpretation of the wave function is an 
intrinsic quantum property of the wave function not depending on
the interactions (which can be understood even classically) with other particles.
We also note that the wave function is complex (not real) even for
uncharged particles, which should be distinguished from fields which are
real for uncharged particles. Essentially, this is because the wave function is defined 
as a superposition of waves with positive frequencies, while the 
corresponding field contains modes with both positive and negative frequencies.
For more details on the relation between field operators and the corresponding
c-number valued wave functions
see, e.g., \cite{ryder,schw,nikfpl1,nikfpl2,nikijmpa,nikmyth}.

All physical superstring states satisfy the Hamiltonian constraint
$(\hat{H}-a)|\Psi\rangle=0$, where $a$ is a constant that depends
on the ordering of the operators. For example, when the normal ordering is used
and open strings are considered, then $a=1/2$ (1) for the NS (R) sector, respectively. 
In the Schr\"odinger picture, the Hamiltonian constraint can be written as
\cite{nikbosfer}
\begin{eqnarray}\label{KGS2}
\left( \int d\sigma \left[
\frac{\eta^{\mu\nu}}{2} 
\frac{\delta}{\delta X^{\mu}(\sigma)} \frac{\delta}{\delta X^{\nu}(\sigma)}
%\right. \right.
%\nonumber \\
%\left. \left. 
- \frac{1}{2} \frac{\partial X^{\mu}(\sigma)}{\partial\sigma} 
\frac{\partial X_{\mu}(\sigma)}{\partial\sigma} 
\right. \right.
\nonumber \\
\left. \left. 
+\hat{\cal H}_{\rm F} \right]  -a \right) \Psi[X,M]=0,
\end{eqnarray}
where
\begin{equation}
\hat{\cal H}_{\rm F}\equiv \frac{i}{2} \eta_{\mu\nu}
\hat{\psi}^{{\sf T}\mu}(\sigma)\rho^0\rho^1\partial_{\sigma}
\hat{\psi}^{\nu}(\sigma) .
\end{equation}
Here we use units $\hbar=c=\alpha'=1$, $\eta^{\mu\nu}$ is the spacetime Minkowski
metric with the signature $(+,-,-,\ldots)$, $\rho^{\alpha}$ are the
matrices satisfying the world-sheet Clifford algebra, $\hat{\psi}^{\nu}(\sigma)$
are operators satisfying the canonical anticommutation relations and carrying
an (unwritten) world-sheet spinor index as well as the spacetime vector
index $\nu$, and $M(\sigma)$ are indices in the infinite-dimensional
Hilbert space in which the operators $\hat{\psi}^{\nu}(\sigma)$ act.
Clearly, (\ref{KGS2}) represents a stringy generalization of the Klein-Gordon
equation. The associated conserved superstring current with a 
consistent probabilistic interpretation is \cite{nikprob}
\begin{eqnarray}\label{strcs}
& J_{\mu}[X;\sigma)= & 
\nonumber \\
&
 i  \displaystyle\int [dM] \, \Psi^*[X,M]
\frac{
\!\stackrel{\leftrightarrow}{\delta}\! }
{\delta X^{\mu}(\sigma)}
\left\{ \prod^{(\sigma)}_{\sigma'}\hat{P}(\sigma') \right\}
\Psi[X,M] ,
&
\end{eqnarray}
where the product $\prod^{(\sigma)}_{\sigma'}$ denotes the
product over all values of $\sigma'$ except $\sigma'=\sigma$, and
\begin{equation}
 \hat{P}(\sigma)=i\left[ 
n^{\mu}(X(\sigma)) \frac{ \stackrel{\rightarrow}\delta } {\delta X^{\mu}(\sigma)} -
\frac{ \stackrel{\leftarrow}{\delta} } {\delta X^{\mu}(\sigma)} n^{\mu}(X(\sigma)) \right] .
\end{equation}
Here $n^{\mu}$ is the unit timelike future-oriented vector normal to the hypersurface
on which the measurement is performed. (In the Bohmian interpretation
$n^{\mu}$ defines a preferred foliation of spacetime \cite{nikprob}, but here we work
in the conventional probabilistic interpretation which does not involve
a preferred foliation.)
The current is conserved in the sense that
\begin{equation}\label{consJ}
\int d\sigma \, \frac{\delta}{\delta X^{\mu}(\sigma)}
J^{\mu}[X;\sigma) = 0.
\end{equation} 
Assuming that the measurement is performed on the hypersurface of constant 
Minkowski time $x^0$, we have $n^{\mu}=(1,0,0,\ldots)$, and
the probability density in the space of string configurations is 
\begin{eqnarray}\label{rhostring}
\rho[X] & = & \left| J_{0}[X;\sigma) \right| 
\\
& = & \left| \int [dM] \, \Psi^*[X,M] 
\left\{ \prod_{\sigma'}\hat{P}(\sigma') \right\}
\Psi[X,M] \right| ,
\nonumber
\end{eqnarray}
where $\prod_{\sigma'}$ denotes the product over all values of $\sigma'$.
For the generalization to an arbitrary hypersurface see \cite{nikprob}.

Another open-superstring constraint having a nontrivial point-particle limit
is $\hat{F}_0|\Psi\rangle=0$ \cite{GSW}, which in the Schr\"odinger picture 
represents a stringy generalization of the Dirac equation \cite{nikbosfer}.
It leads to a different conserved superstring current that generalizes the Dirac current.
However, unlike the Hamiltonian constraint, this constraint is not satisfied by 
all physical states, but only by the states of the R sector.
In analogy with the usual point-particle physics,
one might think that the probability current for the states of the R sector
should be given by this Dirac-like current, while the Klein-Gordon-like current
(\ref{strcs}) should be reserved to the states of the NS sector only.
Nevertheless, in string theory this would not be consistent
with the superposition principle of QM \cite{nikbosfer}, because 
superpositions that contain states from both sectors also exist.
The conclusion is that even the fermionic states from the R sector
must have the probability current defined by the universal 
Klein-Gordon-like current (\ref{strcs}). By similar reasoning,
the universality of the Klein-Gordon-like current (\ref{strcs}) generalizes 
to all closed-string states as well.

Now let us study the point-particle limit. We restrict the analysis to the
case in which $|\Psi\rangle$ is the spin-$\frac{1}{2}$ fermion. 
Therefore, the wave-functional
$\Psi[X,M]$ reduces to the spinor wave function $\psi_m(x)$, where
$m$ is the spinor index \cite{nikbosfer}. The current (\ref{strcs}) reduces to
\begin{equation}\label{curbos}
j_{\mu}(x)=i\psi_m^{*}(x) \!\stackrel{\leftrightarrow\;}{\partial_{\mu}}\! \psi_m(x) ,
\end{equation}
In the usual spinor notation, this can also be written as
\begin{equation}\label{curbo2}
j_{\mu}(x)=i\psi^{\dagger}(x) \!\stackrel{\leftrightarrow\;}{\partial_{\mu}}\! \psi(x) .
\end{equation}
This can also be written in the spinor-covariant form
%\begin{equation}\label{curbo3}
$
j_{\mu}(x)=i\bar{\psi}(x) (\gamma^{\nu}n_{\nu})
\!\stackrel{\leftrightarrow\;}{\partial_{\mu}}\! \psi(x) 
$,
%\end{equation}
where $\bar{\psi}=\psi^{\dagger}\gamma^0$ and $\gamma^{\nu}$
are the Dirac matrices in 10 dimensions.
The probability density (\ref{rhostring}) reduces to the Klein-Gordon probability density
of particle positions \cite{nikijmpa}
\begin{equation}\label{rhoKG}
 \rho_{KG}(x)=|i\psi^{\dagger}(x) \!\stackrel{\leftrightarrow\;}{\partial_{0}}\! \psi(x)| .
\end{equation}
This should be contrasted with the Dirac probability density identified with the time
component of the Dirac current $\bar{\varphi}(x)\gamma^{\mu}\varphi(x)$, i.e.,
\begin{equation}\label{rhoD}
 \rho_{D}(x)=\varphi^{\dagger}(x)\varphi(x) .
\end{equation}
(The spinors $\psi$ and $\varphi$ represent the same physical state normalized according
to the Klein-Gordon and Dirac norm, respectively.
Note also that both the Dirac and the Klein-Gordon current are conserved because
the Dirac spinor satisfies both the Dirac and the Klein-Gordon equation.)

Since we are interested in measurable consequences, in the rest of the discussion 
we restrict the analysis to 4 dimensions and use the notation
$x^{\mu}=(t,x,y,z)$.
The spinor of the electron
moving in the $x$-direction with the momentum $p_x$ and having spin
$+\frac{1}{2}$ is \cite{BD1}
\begin{equation}\label{spinor1}
 \varphi_p=\sqrt{\frac{\omega+m}{2\omega}}
\left( 
\begin{array}{c}
 1 \\ 0 \\ 0 \\
\frac{p_x}{\omega+m}
\end{array}
\right) e^{-ip_{\mu}x^{\mu}} ,
\end{equation}
where $p^{\mu}=(\omega,p_x,0,0)$, 
$\omega=\sqrt{p_x^2+m^2}$, and $m$ is the electron mass.
The normalization is chosen according to the
Dirac-normalization condition $\varphi^{\dagger}\varphi=1$.
The differences between (\ref{rhoKG}) and (\ref{rhoD}) are significant
only for relativistic energies. Therefore, to maximize the difference, 
we study ultrarelativistic electrons with $\omega\gg m$. In this limit
(\ref{spinor1}) simplifies to
\begin{equation}\label{spinor2}
 \varphi_p=\frac{1}{\sqrt{2}}
\left( 
\begin{array}{c}
 1 \\ 0 \\ 0 \\ 1
\end{array}
\right) e^{-i\omega(t-x)} .
\end{equation}
Further, similarly to the spin-0 case \cite{nikprob},
the difference between (\ref{rhoKG}) and (\ref{rhoD})
does not exist for states with definite $\omega$. Therefore,
we study a superposition
\begin{equation}\label{supD}
\varphi=\frac{1}{\sqrt{2}}(\varphi_1+\varphi_2) ,
\end{equation}
where $\varphi_1$, $\varphi_2$ denote spinors (\ref{spinor2})
with frequencies $\omega_1$, $\omega_2$, respectively.

\begin{figure}[t]
\includegraphics[width=6.1cm]{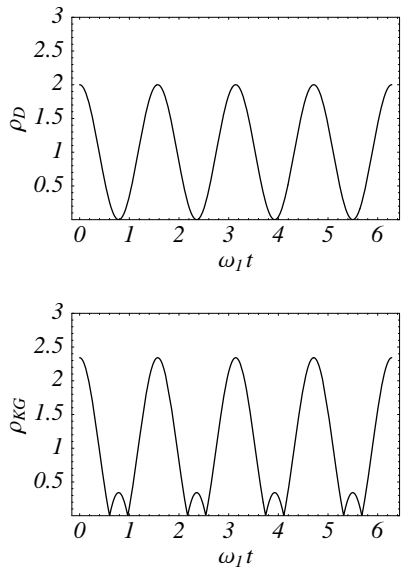}
\caption{\label{fig1}
The probability densities $\rho_{D}$ and  $\rho_{KG}$ as functions of
time, for $\eta=5$.}
\end{figure}

\begin{figure}[t]
\includegraphics[width=6.1cm]{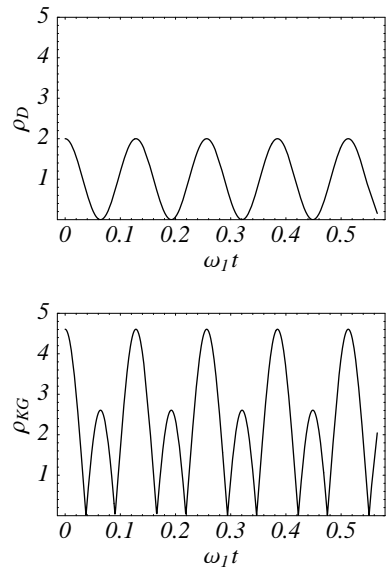}
\caption{\label{fig2}
The same as Fig.~\ref{fig1}, but for $\eta=50$.}
\end{figure}

The superposition (\ref{supD}) normalized according to the 
Klein-Gordon norm is
\begin{equation}\label{supKG}
\psi=\frac{1}{\sqrt{2}} \left( \frac{\varphi_1}{\sqrt{2\omega_1}}
+\frac{\varphi_2}{\sqrt{2\omega_2}} \right) .
\end{equation}
Now the only non-vanishing components of (\ref{curbo2}) are
\begin{equation}\label{j01}
j^0=j^1=1+ 
\frac{\omega_1+\omega_2}{2\sqrt{\omega_1\omega_2}} \, {\rm cos}
[(\omega_1-\omega_2)(t-x)] .
\end{equation}
In general, to obtain the total probability (equal to 1
if an appropriate readjustment of the wave-function normalization is done)
of particle positions,
one should integrate the probability density $\rho_{KG}$ over a hypersurface
chosen such that each integral curve of the vector field $j^{\mu}$
crosses the hypersurface only ones \cite{nikijmpa,nikprob}, which may be non-trivial
when $j^0$ is negative at some regions of spacetime. 
(This global interpretation of $\rho_{KG}$ is the only non-trivial aspect that emerges
from negative values of $j^0$.) However, in our special case in which 
$\varphi_1$ and $\varphi_2$
have the same direction of the ultrarelativistic momentum, we see that $j^0=j^1$, which
implies that all integral curves are straight lightlike lines. Consequently, each Cauchy 
hypersurface of constant $t$ is crossed only ones with each integral curve, implying
that $\rho_{KG}=| j^0|$ represents the probability density {\em everywhere} on any 
Cauchy hypersurface of constant $t$.
 
The state (\ref{supD}) or (\ref{supKG}) describes an electron 
moving in the $x$-direction, prepared in 
a coherent superposition of equally probable frequencies $\omega_1$ and $\omega_2$.
To further specify the measurement setup, we put
the detector of electrons on a fixed position, say $x=0$.
Thus (\ref{j01}) gives
\begin{equation}\label{rhoKGm}
\rho_{KG}(t)=\left| 1+ 
\frac{1+\eta}{2\sqrt{\eta}} \, {\rm cos}[(1-\eta)\omega_1 t] \right| ,
\end{equation}
where $\eta\equiv \omega_2/\omega_1$. 
Without loosing on generality we take $\omega_2>\omega_1$, so $\eta>1$.
For comparison, the corresponding Dirac probability density is simply
\begin{equation}\label{rhoDm}
\rho_{D}(t)=1+ {\rm cos}[(1-\eta)\omega_1 t]  .
\end{equation}
The functions (\ref{rhoDm}) and (\ref{rhoKGm}) are drawn on Fig.~\ref{fig1}
for $\eta=5$ and Fig.~\ref{fig2} for $\eta=50$.
As a consequence of interference between waves with frequencies
$\omega_1$ and $\omega_2$,
the Dirac distribution of electrons as a function of time shows
a usual oscillatory behavior with the frequency $\omega_2-\omega_1$.
The Klein-Gordon distribution shows an unusual oscillatory pattern
containing primary peaks (which correspond to the peaks of 
the Dirac distribution) as well as the secondary peaks (which have no
analog in the Dirac distribution). The primary peaks are allways higher
than the secondary ones, but the relative height of secondary peaks compared to the
primary ones increases with $\eta$. Thus, a possible experimental
verification of the existence of secondary peaks (in addition
to the standard primary peaks) would represent a confirmation
that the probability density of electrons is given by the Klein-Gordon
(rather than Dirac) current, which is a prediction of string 
theory not shared by the usual form of point-particle theory of electrons.

Concerning the practical measurability of the proposed effect, few remarks
are in order. To start, one needs to prepare an electron in a coherent superposition
of two significantly different energies $\omega_1$ and $\omega_2$.
This is certainly not a trivial task, but it might be possible to achieve with 
present technology. A more difficult requirement is the measurement
of the electron position with a precision better than its Compton length
$m^{-1}$, as well as the measurement of the time of detection with
roughly the same precision. Indeed, it is sometimes argued
that a position of a particle cannot be measured with a precision
better than its Compton length. Fortunately, it certainly
cannot be fundamentally true, because it would imply that the
position of the massless photon could not be measured at all, which is
certainly not the case. Nevertheless, it is certainly very difficult to achieve
such a precision in practice and it is not clear if it is possible to achieve this
with present technology. Still, the 
electron Compton length $m^{-1}$ is 22 orders 
of magnitudes larger than the Planck length at which other generic predictions
of string theory become measurable, making our proposal 
much more promising than a direct test of string theory at the Planck scale.

Finally, we note that our prediction of secondary peaks is also interesting
from the point of view of foundations of relativistic QM. 
For example, the existence of the secondary peaks is predicted also for photons 
(the Maxwell equations, of course, also can be viewed as a 
variant of the Klein-Gordon equation), which probably could be tested more easily than for
electrons. Such a test with photons would say nothing about string theory,
but would represent a test of the probabilistic interpretation
of the Klein-Gordon current.

This work was supported by the Ministry of Science of the
Republic of Croatia under Contract No.~098-0982930-2864.

\end{document}